\documentclass[aps,prl,twocolumn,showpacs,preprintnumbers,amsmath]{revtex4}
\usepackage{bm}
\usepackage{amsmath}
\usepackage[dvips]{graphicx}

\begin{document}
\bibliographystyle{apsrev}

\title{\bf Quantum chaos and fluctuations in isolated nuclear spin systems}

\author{J. A. Ludlow and O. P. Sushkov}


\address{School of Physics, University of New South Wales, Sydney 
2052, Australia}

\begin{abstract}
Using numerical simulations we investigate dynamical
quantum chaos in isolated nuclear spin systems. 
We determine the structure of quantum states, investigate the
validity of the Curie law for magnetic susceptibility and find the spectrum of 
magnetic noise. The spectrum is the same for positive and negative temperatures.
The study is motivated by recent interest in condensed-matter experiments 
for searches of fundamental parity- and time-reversal-invariance violations.
In these experiments nuclear spins are cooled down to microkelvin temperatures
and are completely decoupled from their surroundings.
A limitation on statistical sensitivity of the experiments arises from the
magnetic noise.
\end{abstract}


\maketitle

{\it Introduction}.
Much of the present knowledge about violations of the fundamental symmetries P
(invariance under spatial inversion) and T (invariance with respect to
time reversal) comes from experiments measuring P- and T-violating
permanent electric dipole moments (EDM's) of atoms, molecules and the neutron,
see, for example, Ref. \cite{khrip}. Most EDM experiments measure precession
of the angular momentum of the system in an applied electric field analogous 
to the Larmor precession in an applied magnetic field.

In addition to such precession experiments, there are EDM searches of another 
kind \cite{shap,vas}, which have drawn recent renewed attention
\cite{lam,lam1,hunter,sush,DK,budker,KD}.
The idea of these experiments is the following. Suppose that we have some
condensed matter sample with uncompensated spins. 
If an electric field  is applied to the sample, it interacts with the 
associated (P- and T-violating)
EDM's, leading to a slight orientation of the spins in the direction of
the electric field. This orientation, in turn, is measured by measuring the
induced magnetization of the sample.
In this work we concentrate on effects related to nuclear spins in 
insulators with fully compensated electron spins \cite{sush,budker}.
The EDM energy shift under discussion is about $10^{-24}-10^{-28}$eV
per nuclear spin. For comparison a similar shift is created by a magnetic field 
$B \sim 10^{-16}-10^{-20}$T.
So the effect is tiny and limitation to statistical sensitivity comes 
from the fact that the number of spins, $n \sim 10^{23}$, inspite of being
large is still finite \cite{budker}. Basically the limitation comes from a 
kind of magnetic shot noise.
In the EDM experiments the nuclear spins must be cooled down at least 
to $100\mu $K and optimally down to $10-100$nK. At low temperatures the 
spins are completely decoupled from the crystal lattice, so there is no 
contact with any heat bath.
This motivates the problem considered in the present work: magnetic noise
of an isolated nuclear spin system. An important point is that
the total spin of all nuclei is not conserved because magnetic dipole-dipole 
interaction depends on relative orientation of nuclei.

Concerning previous work we first of all refer to the 1959 paper by Hebel and 
Slichter \cite{HS} where the physical meaning of  temperature for an isolated 
spin system has been discussed. This work assumed the validity of a statistical
approach for the isolated quantum system. This is what nowadays is called
dynamical quantum chaos. 
The problem of the onset of dynamical quantum chaos has been addressed much 
later. Level statistics in a spin system with a mobile fermion has been 
investigated by Montambaux et al in 1993 \cite{MPBS}. 
The criterion for onset of quantum chaos in spin glass shards with Heisenberg
interaction in a random external magnetic field has been derived by 
Georgeot and Shepelyansky in 1998  \cite{shep1}.
There have been also  numerical studies of level statistics
in XYZ spin chains with and without a random magnetic field that demonstrated
non-Wigner behavior in the absence of magnetic field \cite{Kudo}.

{\it Model.}
To be specific, we consider the Lead Titanate ferroelectric suggested
for EDM experiments in Refs. \cite{leggett,sush}.
The interaction between the nuclei is governed by the magnetic dipole-dipole 
interaction, the  strength of which falls off as $1/r^3$,
$J_{\alpha\beta}=\gamma^2(\delta_{\alpha\beta}-3n_{\alpha}n_{\beta})/r^3$ \ .
Therefore in the
paramagnetic phase that we are interested in it is sufficient to only
include nearest-neighbour interactions. 
The dominating contribution to the 
dipole-dipole interaction is from the $^{207}$Pb isotope which has a 
natural abundance of 22.1\% and are distributed randomly through 
the lattice. As a consequence the interaction is anisotropic
and random in strength. Therefore we adopt a model Hamiltonian of the form,
\begin{equation}\label{ham}
H=\sum_{\langle kl\rangle}
\sum_{\alpha\beta}J_{\alpha\beta}^{kl}S_{k\alpha}S_{l\beta}
-B\sum_kS_{kz}
\end{equation}  
where $J_{\alpha\beta}^{kl}$ is the interaction between spins $S=1/2$ on
nearest sites $k$ and $l$ on a  lattice and
$B$ is a uniform external magnetic field. The interaction 
$J_{\alpha\beta}^{kl}$ is represented by random numbers uniformly distributed 
between [-J,J]. The typical value of $J$ is
$J \sim 10^{-12}eV \sim 10nK$ \cite{sush}. 
For the sake of simplicity unless otherwise is stated we take a usual square
lattice with periodic boundary conditions.
We will consider the case when the  tensor $J_{\alpha\beta}$ has only 
diagonal components, $\alpha=\beta$, and the case 
when it has both diagonal and off diagonal components. The latter case 
corresponds to real dipole-dipole interaction.
We diagonalize the Hamiltonian (\ref{ham}) numerically exactly and we need 
to know all eigenstates and all eigenenergies, so the size of the matrix
is $2^n$ where $n$ is the number of spins. Therefore, practically we are 
able to consider only relatively small clusters with n=8,10,12. 
For a sufficiently large lattice all results are expected to be self-averaged.
However, we consider relatively small clusters, therefore, to improve
statistics we average results 
over 100 random realizations of $J_{\alpha\beta}^{kl}$.
We do not consider an odd number of spins to avoid Kramers degeneracy of 
spectra that cannot be important in the thermodynamic limit.

{\it Structure of chaotic eigenstates}.
{\underline {Small magnetic field}}.
At zero magnetic field the Hamiltonian (\ref{ham}) is invariant under time 
reversal (T-reversal). This implies that the eigenfunctions of the Hamiltonian 
are split in two sectors of different time-parity: in the first sector they stay 
the same (+) and in the second one they  change sign (-) under the action of all 
the spins being flipped. Therefore the expectation value of magnetization that is 
a T-odd operator vanishes for any state, $\langle i|S_z|i\rangle=0$. 
Because of random $J$, chaos is established in each of these two sectors but
the sectors do not interact.
Anderson localization of single particle states in a random potential is a well 
known effect. In principle a spatial localization of many-body quantum states is also possible,
while it is known that interaction tends to destroy Anderson localization \cite{S}.
Spatial localization would imply Poisson level statistics
(the distribution of level spacing between closest levels), $P_P(s)=\exp(-s)$,
within a sector with a given T-parity in a sufficiently large system. 
On the other hand the Wigner-Dyson distribution $P_{WD}(s)=\frac{\pi s}{2}\exp(-\pi s^2/4)$
within a given sector indicates a full chaotization including delocalization.
To characterise to what degree the statistics of levels are Wigner or 
Poisson, following Ref.~\cite{shep1} we use the parameter 
$\eta=\left.\int_0^{s_0} \left[ P(s)-P_{WD}(s)\right] \rm{d}s\right/
{\int_0^{s_0} \left[ P_P(s)-P_{WD}(s)\right] \rm{d}s}$ \ ,
where $P(s)$ are the statistics measured in numerical simulations and
$s_0=0.4729$ is the intersection point of $P_P(s)$ and $P_{WD}$. 
Localization effects are always enhanced in the 1D case. 
Therefore to investigate the localization scenario we studied level statistics within a given 
T-parity sector for the Hamiltonian (\ref{ham}) on a 1D ring  at $B=0$. We found that 
$\eta=0.18, 0.051, 0.035$ for n=8, 10 and 12 respectively.
So, we do not observe any deviation from the Wigner-Dyson distribution within the accessable
system size. Thus, we come to a conclusion that there is no spatial 
localization of quantum states in the random spin system.
This conclusion is similar to that of Ref. \cite{S} for mobile interacting
particles.

\begin{figure}[h!]
\includegraphics[clip,width=5.0cm]{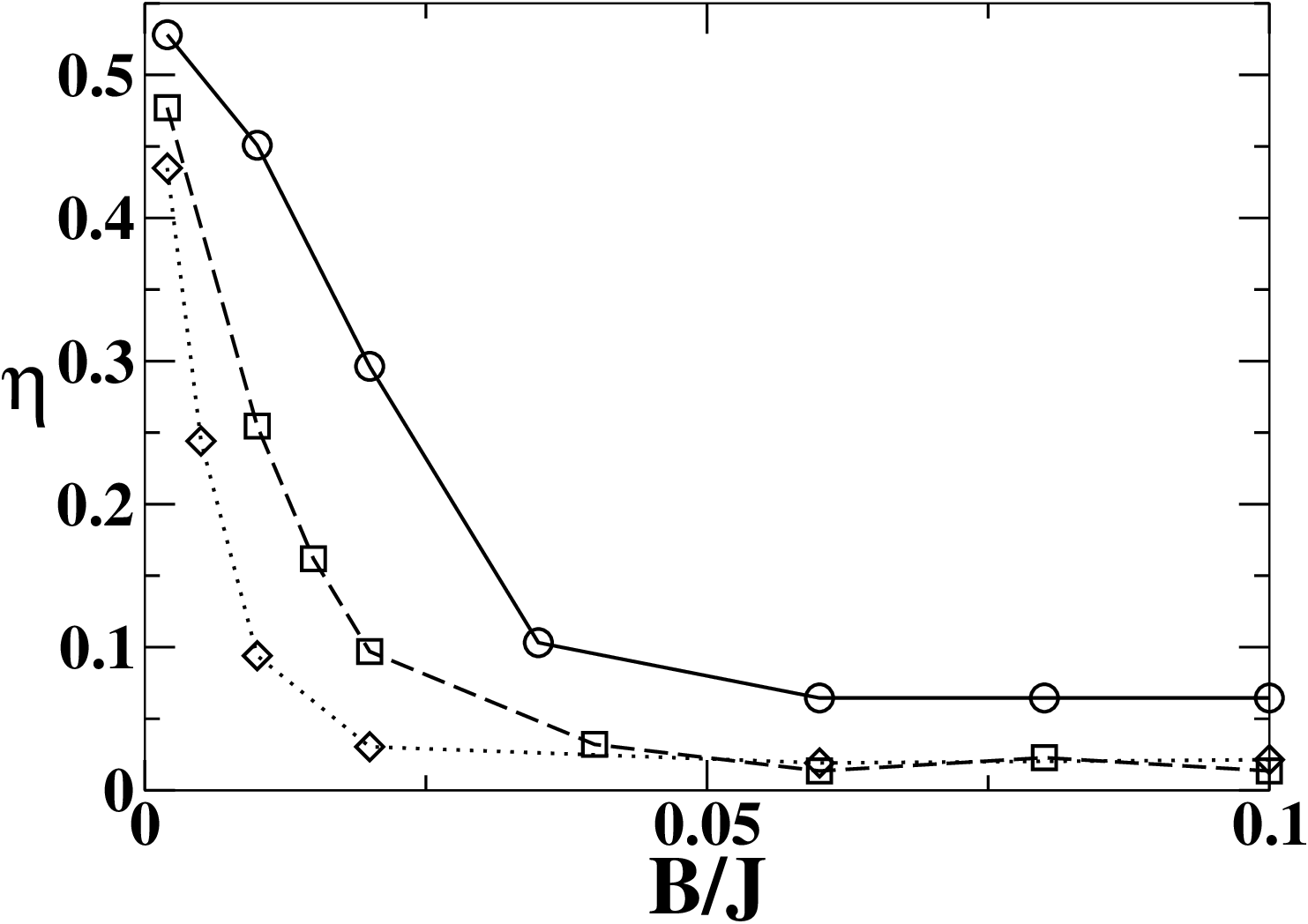}
\vspace{-10pt}
\caption{$\eta$ as a function of the magnetic field: solid line $n=8$,
dashed line $n=10$, dotted line $n=12$.}
\label{thresgr}
\end{figure}

The combined statistics of levels including both sectors at $B=0$ is given by the sum of two 
Wigner distributions, giving an intermediate statistics with $\eta\approx 0.5$, see Fig.\ref{thresgr}.
A very small critical magnetic field is needed to mix the sectors and 
hence to lead to a single  Wigner distribution.
The critical magnetic field $B_{c1}$ is given by the condition that mixing
of two nearest states from different sectors is of the order of unity
\begin{equation}\label{crit}
\frac{B_{c1}\langle j_-|S_z|i_+\rangle}{\Delta E}\sim 1 \ .
\end{equation} 
Here $|i_+\rangle$ and $|j_-\rangle$ are opposite T-parity eigenstates of the 
Hamiltonian (\ref{ham}) at zero magnetic field.
The level spacing is roughly equal to $\Delta E \sim J\sqrt{n}/2^n$, where the factor
$\sqrt{n}$ comes from the total width of the spectrum that is discussed below.
To estimate a typical mixing matrix element in (\ref{crit}) we use the sum rule
$\sum_j \langle i_+|S_z|j_-\rangle \langle j_-|S_z|i_+\rangle=
\langle i_+|S_z^2|i_+\rangle
=\langle i_+|\sum_mS^2_{zm}+\sum_{m\ne k}S_{z,m}S_{z,k}|i_+\rangle
\approx\frac{n}{4} \ .$
Here we take into account that $\sum_mS^2_{zm}=n/4$ and that $\sum_{m\ne k}\to 0$ 
because it is incoherent.
There are $2^{n-1}$ terms in $\sum_j$. Therefore the typical mixing matrix element is
$|\langle j_-|S_z|i_+\rangle|\sim \sqrt{{n}/{2^{n+1}}}$.
Hence substitution in (\ref{crit}) gives the following estimate
\begin{equation}\label{bcrit}
B_{c1}\sim \frac{J}{\sqrt{2^{n-1}}} \ .
\end{equation} 
To confirm this analytical estimate we present in Fig. \ref{thresgr}
the value of $\eta$ calculated numerically for different magnetic fields.
As we already mentioned at zero magnetic field $\eta \approx 0.5$
indicating intermediate statistics, and at large field $\eta \to 0$
indicating the Wigner-Dyson distribution. Taking the value
$\eta =0.25$ as a crossover point we find
$B_c(n+2)/B_c(n) \approx 0.5$ in good agreement with Eq. (\ref{bcrit}). 
Thus the value of the first critical field $B_{c1}$ drops down exponentially
with number of spins. For the EDM experiment we are interested in
large systems, $n\sim 10^{23}$. So in this situation $B_{c1}=0.$
Note that Eq. (\ref{bcrit}) is not a criterion for onset of quantum chaos.
This is the criterion for a crossover from a two component
chaotic distribution to a single component chaotic distribution or in
other words the criterion for destruction of T-parity.
This is quite similar to the dynamical enhancement of spatial
parity violation  in nuclei \cite{SF}.

{\underline {``Strong'' magnetic field}}.
In a strong uniform magnetic field stationary states can be classified by
total spin projection on the direction of the field, $S_z$. 
This regime is realized when the level spacing $B$ is larger than the matrix 
element between directly coupled states that is of the order of $J$, 
see e.~g.~\cite{shep1}.
Thus the crossover to the strong magnetic field regime happens when
$B > B_{c2} \sim J \sim 10^{-7}$T
(the numerical value corresponds to Lead Titanate).
In the strong field regime due to random exchange interaction $J$ the states 
with a given value of $S_z$ are completely mixed up and their energies, according to
our calculations, are spread 
within the band of width $\approx 0.3\sqrt{n}J$. Note that ``conservation'' of total
$S_z$ does not mean that the z-projection of a particular spin is conserved.
Every particular spin fluctuates very strongly. 
Thus the energy spectrum consists of successive bands, separated by $B$,
the width of the band is about $\sim 0.3\sqrt{n}J$. There is the 
Wigner-Dyson statistics of levels within a given band. 
When $n \gg (B/J)^2$ the bands overlap and this leads to 
Poisson level statistics.

The above discussion of the structure of quantum states is valid in the case
when the tensor $J_{\alpha\beta}$ in Hamiltonian (\ref{ham}) has both diagonal and
off diagonal components. In the case without off diagonal components there is
a hidden integral of motion. The Hamiltonian acting on a basis state with 
a particular $S_z$ only changes $S_z$ by $0,\pm 2$, therefore $S_z(mod2)$
is conserved.
This means that eigenfunctions can be expanded in a basis with either even or 
odd $S_z$ components. Then the above considerations are valid separately
for $S_z$-even and for $S_z$-odd sectors.
It is possible that the non-Wigner behaviour observed in Ref.\cite{Kudo} is
due to the hidden integral of motion.

\begin{figure}[h!]
\includegraphics[clip,width=4.0cm]{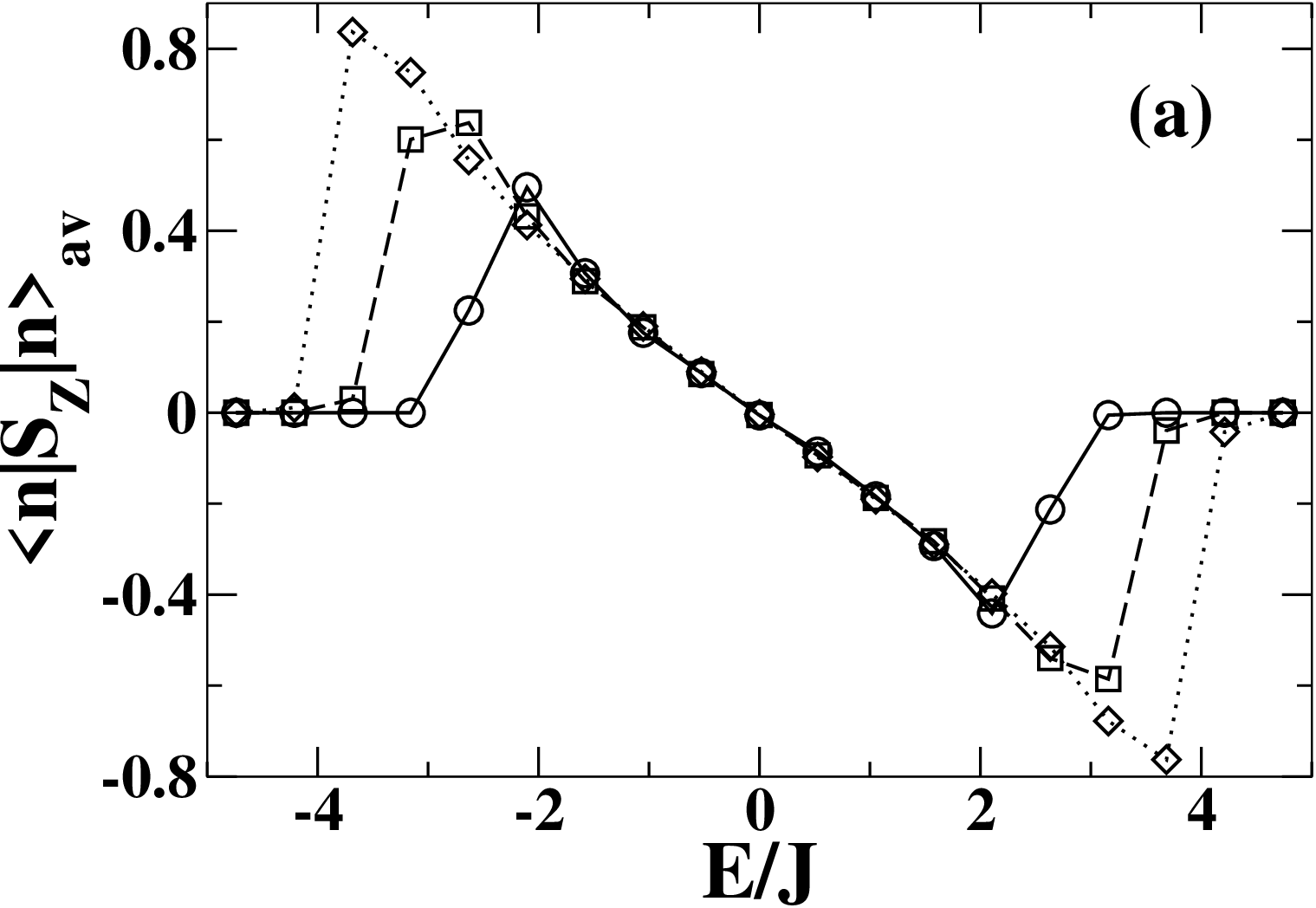}
\includegraphics[clip,width=4.0cm]{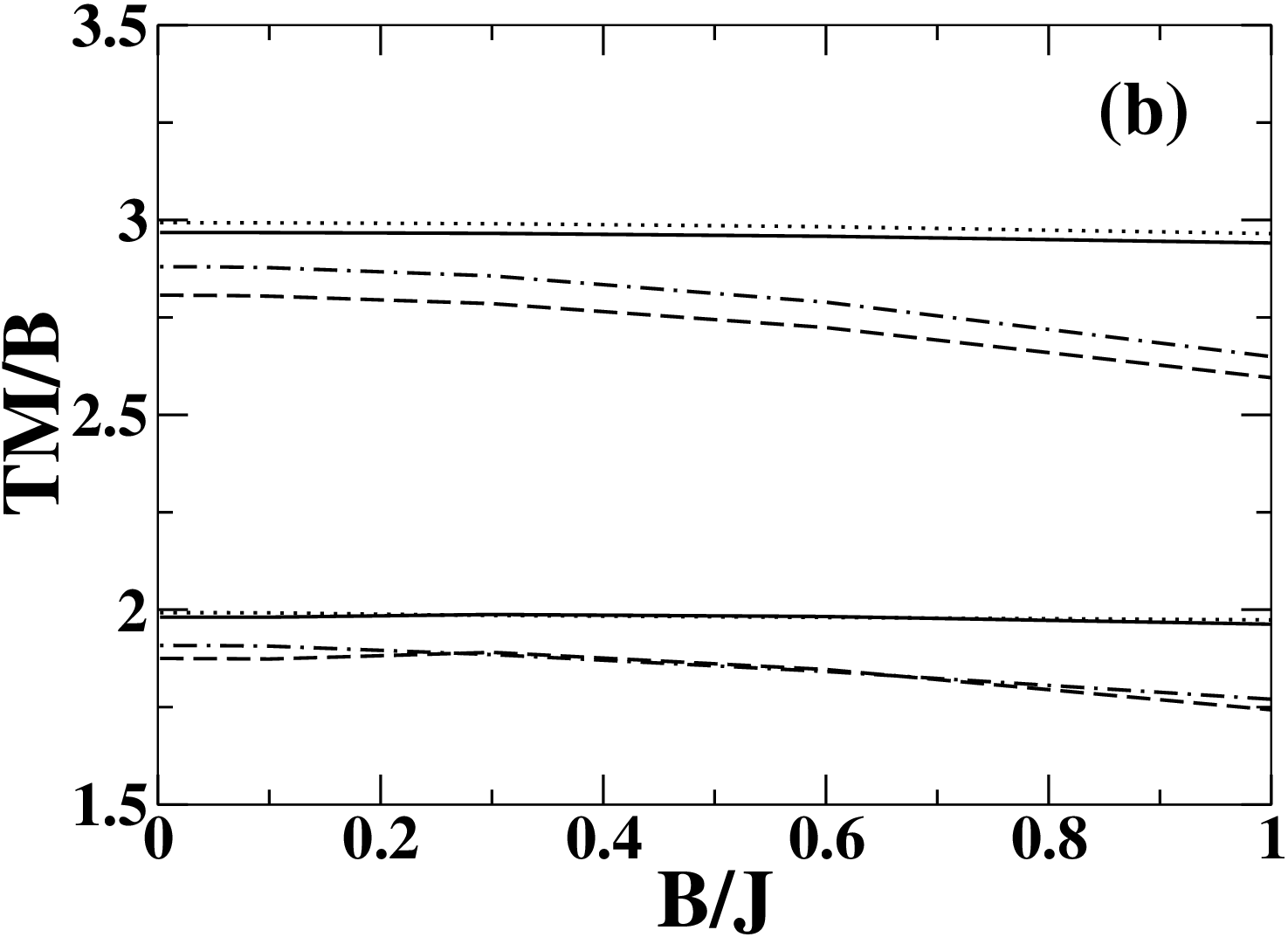}
\vspace{-10pt}
\caption{{\bf a}. Magnetization $\langle i|S_z|i\rangle$ averaged over energy bins $\Delta E=0.5J$.
The magnetic field $B=0.1J$. Solid, dashed, and dotted lines correspond to clusters of 
size $n=8,10,12$ respectively.\\
{\bf b}. The value $TM/B$ versus magnetic field for 
temperatures T=-3J (solid), -J (dashed), J (chain), 3J (dotted)
and for cluster sizes $n=8,12$. The lower group of lines correspond to
$n=8$ and the upper group to $n=12$.
}
\label{engr}
\end{figure}

{\it Temperature, average energy and  magnetic susceptibility}.
The textbook analysis of a spin system based on assuming the validity
of the Boltzmann distribution gives the following well known relations \cite{land}
\begin{eqnarray}
\label{chi}
\chi=\frac{n}{4T} \ , \ \ \ \
E(T)={\overline {E}}_i-\frac{1}{T}\left({\overline {E_i^2}}-
{\overline {E}}_i^2\right) \ .
\end{eqnarray}
Here $\chi$ is the magnetic susceptibility, $E(T)$ is the average energy
at a given temperature (we set the Boltzmann constant equal to unity), 
${\overline {E}}_i$ is the average energy of stationary states and 
${\overline {E_i^2}}$ is the average energy squared.
${\overline {E}}_i$ and ${\overline {E_i^2}}$ are
independent of temperature.
First we want to check the validity of Eq.(\ref{chi}) for the isolated dynamical system
(\ref{ham}). In this case Eq.(\ref{chi}) in essence defines an effective temperature,
see the discussion in Ref.\cite{HS}.
To give a precise meaning to Eq. (\ref{chi}) one has to consider many energy levels 
inside a bin around some given energy $E$. Then, according to (\ref{chi}) the
 susceptibility  averaged over these levels is related to the energy $E$.
We have checked numerically that for the Hamiltonian (\ref{ham}) ${\overline {E}}_i\approx 0$
and ${\overline {E}}_i^2\approx 0.13J^2n$. Hence, according to (\ref{chi}) $\chi \approx -1.94 E/J^2$.
This agrees well with results of numerical simulations with Hamiltonian (\ref{ham})
shown in Fig. \ref{engr}a.
This figure represents values of the level magnetization  $\langle i|S_z|i\rangle$ 
calculated in the field $B=0.1J$, averaged over bins of width $\Delta E=0.5J$,
and averaged over random $J_{\alpha\beta}$. Deviation from linear dependence at large positive 
and negative energy is the finite size 
effect: Eqs. (\ref{chi}) make sense only if the average energy is smaller than the
``band width'',  $|E|\lesssim \sqrt{n}J$.
It is instructive to consider instead of ``rectangular binning''
another way of sampling of energy levels.
In particular the  Boltzmann sampling defined as
$M=\frac{1}{Z}\sum_i \langle i|S_z|i\rangle e^{-E_i/T}$,
where $Z=\sum_{i}e^{-E_i/T}$.
Let us stress that temperature here is just a parameter that samples a set of 
energy levels around the average energy that according to (\ref{chi})
is $E=-0.13nJ^2/T$. One can also check that the energy ``window'' around the
average energy is $\delta E_{rms}=\sqrt{0.13n}J$.
Note that the Boltzmann sampling does not imply a validity of the equilibrium
Boltzmann distribution. For example we can use Boltzmann sampling in a system
without interaction that never comes to an equilibrium.
The value of $TM/B$ calculated for different cluster sizes, temperatures 
(positive and negative) and magnetic fields and averaged over  
random $J_{\alpha\beta}$ is presented in Fig. \ref{engr}b. For high temperatures, $|T|=3J$, 
the results agree perfectly with the Curie law. Unexceptedly it agrees even 
at $B \lesssim B_{c1}$. For lower temperatures, $|T|\sim J$ the results deviate from 
Curie's law as one should expect.
To complete the analysis we have also calculated the dispersion of the susceptibility 
and found that
$\delta \chi=\sqrt{\langle\chi^2\rangle-\langle\chi\rangle^2}
\sim 0.01\sqrt{n}/T$.

\begin{figure}[h!]
\includegraphics[clip,width=4.5cm]{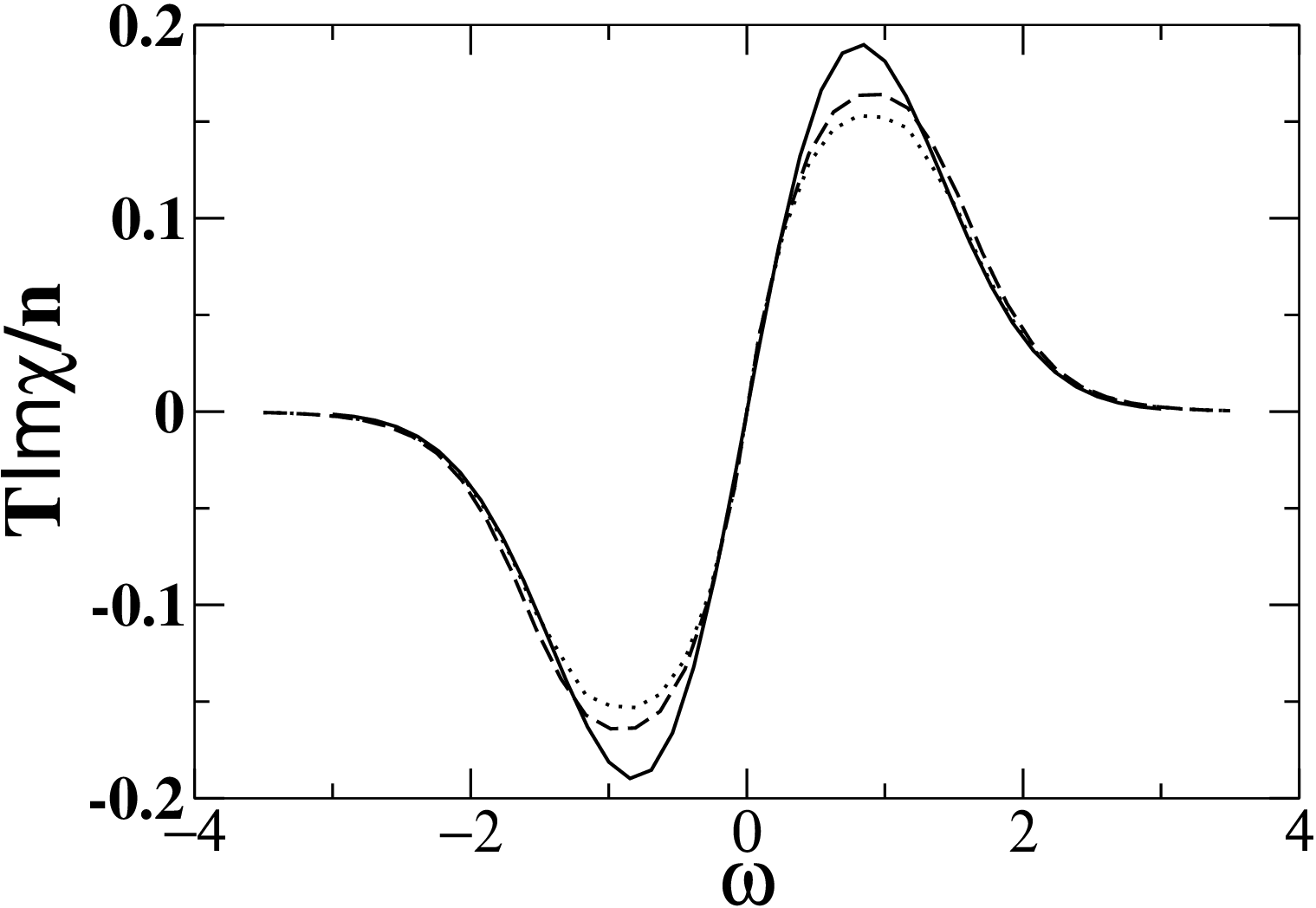}
\vspace{-10pt}
\caption{$TIm\chi /n$ versus $\omega$. 
Solid line, analytic. Numerical results for $n=12$
at  $T=3$ and $T=1$ are shown by dashed and dotted lines respectively.}
\label{gengraph}
\end{figure}

{\it Magnetic susceptibility at nonzero frequency, fluctuations}.
There are no temporal fluctuations in an isolated system in a given quantum 
state.
The fluctuations come from the fact that a large system cannot be prepared in a fixed
quantum state. We always deal with a density matrix that represents a 
combination of a large number of quantum states around some average energy.
In this section we consider only the Boltzmann sampling:  the average energy is 
$E=-0.13nJ^2/T$ and the 
energy ``window'' is $\delta E_{rms}=\sqrt{0.13n}J$.
In this case we can use the standard formula for the imaginary part of the magnetic susceptibility
\cite{land}
\begin{eqnarray}\label{imsus}
Im[\chi(\omega)]&=&
\frac{\pi}{\hbar}\left(1-e^{-\hbar\omega/T}\right)
\frac{1}{Z}\sum_{i,j}e^{-E_i/T}\nonumber\\
&\times&|\langle i|S_z|j\rangle|^2\delta (\omega + \omega_{ij}) \ .
\end{eqnarray}
Here $\omega_{ij}=(E_i-E_j)/\hbar$.
First we perform direct numerical simulations using this formula and quantum 
states generated by the Hamiltonian (\ref{ham}) at $B=0.1J$ \cite{com1}
For simplicity we consider the case when the tensor $J_{\alpha\beta}$ 
has only diagonal components. Results of simulations for cluster $n=12$
are shown in Fig.~\ref{gengraph}. The plots for $n=8,10$ are very similar.
The result unambiguously indicates a Gaussian line shape, 
$Im[\chi]\approx \frac{n}{4T}\sqrt{\pi}\omega\tau e^{-\omega^2\tau^2} \ ,$
with $\tau\approx 0.85/J$. Assuming the Gaussian shape we can also calculate
$\tau$ analytically using a method usually applied in NMR studies \cite{slichter}.
At a large temperature, $E_i \ll T$, and hence we can represent (\ref{imsus}) as 
$Im[\chi(\omega)]=\frac{\pi\omega}{T}f(\omega)$
where the lineshape $f(\omega)$ is given by,
$f(\omega)=\sum_{i,j}
|\langle i|S_z|j\rangle|^2\delta (\omega + \omega_{ij})$.
Let us calculate the second moment of $f(\omega)$,
$M_2=\left.\int_{-\infty}^{\infty}\omega^2 f(\omega){\rm d}\omega\right/
\int_{-\infty}^{\infty}f(\omega){\rm d}\omega$.
Using the completeness relation and the spin decoupling $S_z^2=\sum_{kl}S_{zk}S_{zl}\to
1/4\sum_{kl}\delta_{kl}$ (the indexes $k$, $l$ enumerate sites) we find the denominator
$\int_{-\infty}^{\infty}f(\omega){\rm d}\omega=n2^{n-2}$.
To calculate the numerator one represents it in terms of the commutator of spin with the
Hamiltonian 
$\int_{-\infty}^{\infty}\omega^2f(\omega){\rm d}\omega= \sum_i \langle i|[H,S_{z}]^2|i\rangle$.
Then calculating the commutator and  using the spin decoupling we find
\begin{eqnarray}\label{intf2}
&&\int_{-\infty}^{\infty}\omega^2f(\omega){\rm d}\omega=\frac{2^n}{12}
\sum_{\langle kl\rangle} \left[(J_{xx}^2+J_{yy}^2+J_{zz}^2)\right.\\
&&-(J_{xx}J_{yy}+J_{yy}J_{zz}+J_{xx}J_{zz})
+\left.3(J_{xy}^2+J_{zy}^2+J_{xz}^2)\right] \ .\nonumber
\end{eqnarray}
Here $\langle kl\rangle$ represents a pair of nearest sites.
Since we have performed numerical simulations without off diagonal components of $J_{\alpha\beta}$,
the right hand side of Eq. (\ref{intf2}) is $n2^{n-1}J^2/3$ and hence $M_2=2/3J^2$.
On the other hand for the Gaussian line shape $M_2=1/(2\tau^2)$. Hence the analytical calculation
gives the $\tau=\sqrt{3/4}/J$. The Gaussian curve  with this value of the relaxation time
is shown in Fig.~\ref{gengraph} by solid line. It is in good agreement with results of direct
numerical simulations. The real part of the susceptibility can be easily found using Kramers-Kronig
relations (see, for example, Ref. \cite{land})

As soon as $Im[\chi]$ is known then using the fluctuation dissipation theorem 
one can find magnetic noise, i.~e. the spectral density $(M^2)_{\omega}$ of the square of the 
deviation of the  magnetization from its average value  \cite{budker}, $V^2(M^2)_{\omega}=
\hbar \coth(\hbar\omega/2T)Im[\chi(\omega)]$.
$V$ is volume of the sample.
A very interesting point is that at $\omega \ll T$
the noise is independent of temperature and moreover
it is the same for positive and negative temperature.
What is usually called the  fluctuation dissipation theorem in the case of negative temperature
becomes the fluctuation {\it radiation} theorem because $Im[\chi(\omega)]$ changes sign.

We have investigated the structure of chaotic quantum states
in a spin lattice system with random interactions. We also checked
the validity of the Curie law for magnetic susceptibility and find  
the spectrum of magnetic noise. 
The temperature independent noise limits the statistical sensitivity of 
experiments on parity and time-reversal-invariance violations.

We thank D.~Budker. J.~Imry, A.~Luscher, J.~Oitmaa, and A.~Sushkov,  
for stimulating discussions and critical comments. 
This work was supported by the Australian Research Council.


\begin{thebibliography}{99}
\bibitem{com1}
For simulations we replace the $\delta$-function by a smooth function
$\delta(x)\to \frac{1}{\sqrt{2\pi}\sigma} \exp\left(-x^2/2\sigma^2\right)$
where $\sigma=J/2^{n-2}$.
\end{thebibliography}
\end{document}